\documentstyle[12pt]{article}
\font\titolo=cmbx10 scaled\magstep2

\font\tsnorm=cmr12

\font\tscorsp=cmti10

\def\NPB{{\em Nucl. Phys. }}
\def\PLB{{\em Phys. Lett. }}  
\def\PRL{{\em Phys. Rev. Lett. }}
\def\PRD{{\em Phys. Rev. }} 
 
\def\MPLA{{\em Mod. Phys. Lett. }} 
 
\def\PRP{{\em Phys. Rep. }}
\def\IJMP{{ \em Int. J. Mod. Phys. }}
\def\z{Z\kern -4.6pt Z}
\def\dx{\int d^4x\ \sqrt{-g}\ }
\def\to{\rightarrow}
\def\lg{\left\{}
\def\lq{\left[}
\def\lt{\left(}
\def\rg{\right\}}
\def\rq{\right]}
\def\rt{\right)}
\def\lra{\leftrightarrow}
\def\ha{{1\over 2}}
\def\a{\alpha}

\def\e{\eta}
\def\ef{e^{-\eta}}

\def\r{\rho}
\def\s{\sigma}
\def \S{\Sigma}
\def\pu{\psi_{1}}
\def\pt{\psi_{3}}
\def\xd{\chi_{2}}
\def\xq{\chi_{4}}
\def\t{\tau}

\def\be{\begin{equation}}
\def\ee{\end{equation}}
\def\bea{\begin{eqnarray}}
\def\eea{\end{eqnarray}}
\def\bc{\begin{displaymath}}
\def\ec{\end{displaymath}}
\def\lb{\label}
\def\q{\quad}

\begin{document}
\pagestyle{empty}
\null
\vskip 5truemm
\begin{flushright}
INFNCA-TH9719\\
November 1997 
\end{flushright}
\vskip 15truemm
\begin{center}
{\titolo DUALITIES, COMPOSITENESS AND SPACETIME }
\end{center}
\begin{center}
{\titolo STRUCTURE OF 4D EXTREME STRINGY BLACK HOLES  }
\end{center}
\vskip 15truemm
\begin{center}
{\tsnorm Mariano Cadoni}
\end{center}
\begin{center}
{\tscorsp Dipartimento di Scienze Fisiche,  
Universit\`a  di Cagliari,}
\end{center}
\begin{center}
{\tscorsp Via Ospedale 72, I-09100 Cagliari, Italy,}
\end{center}
\begin{center}
{\tscorsp and}
\end{center}
\begin{center}
{\tscorsp  INFN, Sezione di Cagliari.}
\end{center}
\vskip 15truemm
\begin{abstract}
\noindent
We study the BPS black hole solutions  of  the (truncated)  action
for heterotic string theory compactified on a six-torus. 
The $O(3,Z)$ duality symmetry of the theory, together with the bound state 
interpretation of extreme black holes, is used to generate the whole 
spectrum of the solutions. The corresponding spacetime structures, written 
in terms of the string metric,  are 
analyzed in detail. In particular,  we  show that only the elementary 
solutions present naked singularities. The bound states have either 
null singularities (electric solutions) or are regular (magnetic or 
dyonic  solutions) with near-horizon geometries  given by the product 
of two 2d 
spaces of constant curvature. The behavior of some of these solutions 
as supersymmetric attractors is discussed. We also show that our 
approach is very useful to understand 
some of the puzzling 
features of  charged black hole solutions in string theory.
\end{abstract}
\vfill
\begin{flushleft}
\end{flushleft}
\smallskip
\vfill
\hrule
\begin{flushleft}
{E-Mail: CADONI@CA.INFN.IT\hfill}
\end{flushleft}
\vfill
\eject
\pagenumbering{arabic}
\pagestyle{plain}
\section{Introduction}
\paragraph{}
The recent developments in string theory have pointed out the 
crucial role played by non-perturbative structures such as BPS 
solitons, black holes, D-branes, etc. \cite {RD}.  Though we are still 
far away 
from having full control of the non-perturbative regime of string 
theory, the  investigation of these structures represents a way to 
check new ideas, for example the dualities between different 
string theories, and to try to address in a new setting  old, still 
unsolved, problems of black hole physics.

From this perspective, the fact that for the first time the black 
hole entropy has been calculated by counting microstates is very 
encouraging \cite {ST}. Moreover, the new scenario
represent an unifying framework, in which most of the known solutions 
of low-energy string theory find their natural place. This is 
achieved, for instance, by relating 
typically string structures, e.g. D-branes \cite {DB},
with  more conventional spacetime structures, e.g. black  holes.
 
One  important issue, which  has been widely 
investigated, is the interpretation  of the geometrical structures, 
appearing in the low-energy theory, in terms of the fundamental string 
dynamics \cite {DR,PT,TS}. Generically, one expects the elementary  
(solitonic) string excitations
to be related with electric (magnetic or dyonic) 
charged extreme black hole solutions of the 4d low-energy string 
theory.
This correspondence should be particularly 
instructive because  extreme  black holes
are   BPS saturated states,  so that one is tempted   to use 
supersymmetry as a cosmic censor \cite {KLOP}. 

The implementation of the previous idea was only partially 
successful. For example, whereas  extremal dilatonic 
black holes with dilaton coupling  $a= 0,1$ are regular or
have only null singularities, the $a=\sqrt 3$ solution has a naked 
singularity. Supersymmetry alone  fails as universal cosmic censor.
In the context of $N\ge2$ supergravity theories, the requirement that 
the solutions preserve some of the supersymmetries puts 
strong constraints on  the form   of the admissible spacetime structures. 
The fact that the Bertotti-Robinson geometry behaves as an attractor 
for a wide class of $N=2$ supersymmetric black holes is an example of 
this highly constrained dynamics \cite {FK,FK1,FKS}. Moreover, it gives us 
a general principle to compute the black hole entropy as an extremum 
of the central charge \cite {FK}. 

Despite the relevance of these results, some of the open questions that 
have been with us since the early investigations on charged  black holes in 
string theory, remain still unanswered.
Why is the spacetime structure of the extreme $a=0$ solution (the 
Reissner-Nordstrom solution of general relativity) so drastically 
different from that of the other supersymmetric, $a=1,\sqrt 3, 
1/\sqrt 3$ solutions ?
Why does the former have non-vanishing, charge-dependent, entropy,
whereas the others  have zero entropy? Why does the  
near-horizon geometry of the former  behave as an attractor  
whereas this is not the case for the others?

A possible starting point for trying to answer these questions is to 
assume that the different behavior of  the  solutions of the 
low-energy theory can be explained 
in terms of the fundamental string dynamics. A first step in this 
direction is  the interpretation of the black hole solutions
in terms of string states. Progress along this line has been achieved 
by considering the $a=\sqrt3$ solution as elementary and the other 
supersymmetric solutions as bound states \cite {DR,RA}. 
This idea was further 
supported by the 
interpretation of the black holes as intersections of D-branes \cite{PT}.
    
In this paper we investigate, using as guideline the idea of 
compositeness together with that of duality symmetry, the extreme 
black hole solutions of
a truncated $d=4$  low-energy string action. The model is particularly 
interesting 
because it gives an unified  description of $d=4$ 
black hole solutions in string theory. We show that the model has an
$O(3,Z)$ duality symmetry that is generated by the duality (or 
mirror) symmetries connecting different string theories (heterotic, 
$IIA$, $IIB$) and by $S-T-U$ dualities.
We use the $O(3,Z)$ symmetry and the bound state interpretation 
of extremal black holes to generate 
the BPS spectrum of the theory. We try to explain the features of the 
solutions by considering the correspondence with string states.
The  geometrical 
structures  of the solutions are given in terms of the string metric. 
We  show 
that whereas the elementary solutions have naked singularities, the 
bound states correspond to spacetimes with null singularities, in the 
electric, case and to regular spacetimes for the magnetic or dyonic 
solutions.

The structure of the paper is as follows.
In sect. 2  we present  the model we are going to 
investigate and its $O(3,Z)$ duality symmetry. In sect. 3 we consider 
the BPS black hole  solutions  and we  explain
how $O(3,Z)$ can be used as a spectrum-generating symmetry.
The elementary solutions and the rules for constructing bound states 
are presented in sect. 4. In sect. 5 we construct the bound states 
of elementary solutions. In sect. 6 we discuss the corresponding 
spacetime structures. In sect. 7 we investigate  the behavior  
 of the near-horizon solutions   as  supersymmetric 
attractors. We also   use  a minimization procedure to calculate 
the entropy. Finally, in sect. 8 we present our conclusions.
  
\section{ Truncated $d=4$ string action}
\paragraph{}

The starting point of our discussion is the truncated version of   
the bosonic action for heterotic string theory compactified on a 
six-torus \cite{MS,SE},
\bea
 S_{H}&=&{1\over 16\pi G}\dx \lg R-\ha \left[\left( \partial \e\right)^2+
\left( \partial \s\right)^2+\left( \partial 
\r\right)^2\rq\right.\nonumber\\
&-& \left.{\ef\over 4}\lq e^{-\s-\r}F_{1}^{2}+e^{-\s+\r}F_{2}^{2}+
e^{\s+\r}F_{3}^{2}
+ e^{\s-\r}F_{4}^{2}\rq\rg.
\lb{e1}
\eea
In the bosonic action of heterotic string theory 
toroidally compactified to $d=4$, we have set to zero the axion fields 
and all the $U(1)$ field strengths but four, two Kaluza-Klein fields
$F_{1}, F_{2}$ and  two winding modes  $F_{3},  
F_{4}$. The scalar fields are related to the standard definitions of 
the string coupling, K\"ahler form and complex structure of the torus,
through  the equations
\be
e^{-\e}={\rm Im}S,\quad e^{-\s}={\rm Im}T,\quad e^{-\r}={\rm Im}U.
\lb{e2}
\ee
Owing to the $SL(2,Z)_{S}\times O(6,22,Z)_{TU}$ duality symmetry of the 
heterotic string in $d=4$, the truncated theory (\ref{e1}) should have as 
duality group   $ Z_{2}\times O(2,Z)$, where $Z_{2} \subset SL(2,Z)_{S}$ 
is the  strong/weak coupling duality and $O(2,Z)\subset O(6,22,Z)_{TU}$ is a 
symmetry of the action (\ref{e1}), not only of its equations of motion.
However, one can easily verify that the equations of motion following 
from the action (\ref{e1}) are invariant under the bigger group $O(3,Z)$.
This enhancement of the duality symmetry is a consequence of the 
string/string/string triality discussed in ref. \cite{DLR}, which relates 
one with the other the heterotic, type $IIA$ and type $IIB$ strings in 
four dimensions.
From this point of view,  the action (\ref{e1}) can be thought of as a 
truncated, 
$S-T-U$ symmetric, action that gives a unified description of the low-energy 
solutions of string theory. 

In the next sections we will make a broad use of the $O(3,Z)$ duality 
symmetry of the model (\ref{e1}). In particular, we will use it 
to generate  the BPS spectrum and to discuss the features of the 
corresponding spacetime structures. For this reason, we will first 
present some  general properties of this group, its realization in terms of 
the fields appearing in the action (\ref{e1}) and  its relationship with the 
string/string/string triality of ref. \cite{DLR}. 

The $O(3,Z)$ group has 
48  
elements and can be  realized as rotations (with integer entries) 
of the space
$(\e,\s,\r)$, transforming, additionally, 
the field strengths
$F_{i}$ ($i=1..4$). The transformations of the $O(2,Z)_{TU}$ subgroup are 
off-shell symmetries of the theory (they correspond to mirror 
symmetries of string theory), whereas the other transformations are 
on-shell symmetries. An useful way to describe the $O(3,Z)$ group 
is to generate it by using  the three $S-T-U$ duality transformations 
$\t_{S},\t_{T},\t_{U}$ together with the permutation group $P_{3}$ 
acting on the scalar fields  $\e,\s,\r$:

\bea\lb{e3}
\t_{S}&:& \e\to -\e,\q F_{1}\to \tilde {F}_{3},\q 
F_{3}\to \tilde F_{1},\q F_{2}\to \tilde F_{4},\q
F_{4}\to \tilde F_{2};\nonumber\\ 
\t_{T}&:& \s\to -\s,\q F_{1} \leftrightarrow F_{4},\q
 F_{2}\leftrightarrow F_{3};\nonumber\\
\t_{U}&:& \r\to -\r,\q F_{1}\leftrightarrow F_{2},\q 
F_{3}\leftrightarrow F_{4};
\eea
\bea\lb{e4}
P_{3}: & &\s \leftrightarrow  \r,\q  F_{2}\leftrightarrow F_{4};
\nonumber\\
& &\e \leftrightarrow  \s,\q  F_{3}\to \tilde F_{4},\q 
F_{4}\to \tilde F_{3};
\nonumber\\
& &\e\to\s,\q\s\to \r,\q \r\to\e,\q F_{4}\to F_{2}, \q
F_{2}\to \tilde F_{3},\q
F_{3}\to \tilde F_{4};
\nonumber\\
& &\e \leftrightarrow  \r, \q  F_{2}\to \tilde F_{3},\q 
F_{3}\to \tilde F_{2};
\nonumber\\
& &\e\to\r,\q\s\to \e,\q \r\to\s, \q F_{2}\to F_{4},\q 
F_{3}\to \tilde F_{2},\q
F_{4}\to \tilde F_{3};
\eea
 where $\tilde{F}_{1}= {e^{-\e-\s-\r}}{^*}F_{1}$, $\tilde{F}_{2}= 
 {e^{-\e-\s+\r}}{^*}F_{2}$, $\tilde{F}_{3}= {e^{-\e+\s+\r}}{^*}F_{3}$, 
 $\tilde{F}_{4}= {e^{-\e+\s-\r}}{^*}F_{4}$
 ($^{*}$ denotes the Hodge dual). 

The previous transformations generate the 
whole $O(3,Z)$ group. Each element of $O(3,Z)$ can be written as a 
product of transformations appearing in (\ref{e3}),(\ref{e4}). Moreover, 
these 
group-generating transformations have a simple interpretation in 
terms of string duality transformations. $\t_{S}$, $\t_{T}$, $\t_{U}$ 
are the well-known  $S-T-U$ string dualities that act on the moduli 
and on the charges. 
$\t_{S}$  interchanges  electrically (magnetically) charged Kaluza-Klein 
(KK) states 
with magnetically (electrically) charged Winding (W) states. 
$\t_{T}$ maps electrically (magnetically) charged  KK states into
electrically (magnetically) charged  W state. 
$\t_{U}$ maps   electrically (magnetically) charged  KK (W) states into
electrically (magnetically) charged  KK (W) states. 

The transformations (\ref {e4}),
represent string/string dualities (or mirror symmetries). 
The permutation group $P_{3}$ generates the string/string/string triality 
diagram of ref. \cite{DLR}. The five transformations (\ref{e4}) are, 
respectively, the five $d=4$ string/string low-energy  dualities (or mirror 
symmetries): $H_{STU}\to H_{SUT}$, $H_{STU}\to (IIA)_{TSU}$, 
$H_{STU}\to (IIA)_{UST}$, $H_{STU}\to (IIB)_{UTS}$, $H_{STU}\to 
(IIB)_{TUS}$,
where $H, IIA, IIB$ are the three string types 
(heterotic, $IIA, IIB$).

Besides the duality symmetries, an other important feature of the 
model under consideration is the possibility to describe some of its 
 solutions
 as the solutions of a single scalar,
single gauge field action \cite{LP}. In fact, it has been shown that, for 
particular values of the integration constants (the charges and the 
asymptotic values of the scalar fields) the solutions can be 
obtained from a 
single scalar, single gauge field, action with dilaton coupling parameter 
$a=\sqrt{3},1,1/\sqrt{3},0$ \cite {RA}. This fact has been used to give 
a bound 
state interpretation of extremal black holes in string theory. We will 
come back to this point in sects. 4 and 5, when we will generate the
BPS spectrum using the idea of compositeness.

Let us conclude this section with some remarks about the 
supersymmetries of the model we are considering. Generically,  
compactification of the heterotic string on a six-torus produces   $N=4$ 
supergravity in four dimensions. However,  we can also embed the 
solutions
of the action (\ref{e1}) into  $N=8$ supergravity or, equivalently, we can 
consider eq. (\ref{e1}) as a truncated (bosonic) action of $N=8$ 
supergravity. In the following, we will consider this second 
option,  which give rise to four central charges $Z_{i}$.

\section{BPS states and $O(3,Z)$ as spectrum-generating symmetry}
\paragraph{}

The solutions of the action (\ref{e1}) describing extreme black holes 
are by now well known \cite{CY,FK}. We will consider here only 
supersymmetric solutions, i.e. BPS states that saturate some of the 
Bogomol'nyi bounds of $N=8$ supergravity.
We want to analyze the action of the duality symmetries on the space of 
the solutions. In general, duality transformations change the constant 
values 
of the scalar fields  at infinity, which will be denoted here as
$g=e^{\e_{0}/2},h=e^{\s_{0}/2}, l=e^{\r_{0}/2}$ . The form of the solution 
given in ref. \cite{FK} is  the most suitable one for our purposes, because 
it makes the dependence of the solutions on the moduli $g,h$ and $l$ 
manifest,
\bea\lb{e5}
ds^{2}&=& - e^{2U}dt^{2}+e^{-2U} (dr^2+r^2d\Omega^2),\qquad e^{4U}=\pu \pt 
\xd \xq,
\nonumber\\
e^{2\e}&=&{\pu\pt\over \xd\xq},\quad e^{2\s}={\pu\xq\over \xd\pt},\quad   
e^{2\r}={\pu\xd\over \pt\xq},\nonumber\\
F_{1}&=&  d\pu\wedge dt,\quad
\tilde F_{2}=  d\xd \wedge dt, \quad
\quad F_{3}=  d\pt\wedge dt, \quad
\tilde F_{4}=  d\xq \wedge dt, 
\eea
where $\pu,\pt,\xd,\xq$ are given in terms of harmonic functions,
\bea\lb{e6}
\pu&=&\lt {1\over ghl} + {q_{1}\over r}\rt^{-1},\qquad
\pt=\lt {hl\over g} + {q_{3}\over r}\rt^{-1},\nonumber\\
\xd&=&\lt {gh\over l} + {p_{2}\over r}\rt^{-1},\qquad
\xq=\lt {gl\over h} + {p_{4}\over r}\rt^{-1}.
\eea
$q$ and $p$ are, respectively, electric  and magnetic charges. They will 
be taken 
as positive  in order to have only supersymmetric solutions 
(negative charges can give rise to extreme non-supersymmetric black hole
solutions \cite{OR}).
The form of the solution (\ref{e5}) does not give full account  of 
the BPS black hole solutions of the theory. There 
are, for instance, other solutions that are related to those 
appearing in eq.(\ref{e5}) 
by means of duality transformations (\ref{e3}), (\ref{e4}). 

Our goal is to 
generate the BPS 
spectrum by using the $O(3,Z)$ duality symmetry. To do this, we need 
an efficient way to classify the solutions in terms of the 
parameters appearing in eqs. (\ref{e5}),(\ref{e6}).
The generic BPS saturated state will be labeled by the $U(1)$-charge vectors 
$Q_{i},P_{i} \,(i=1..4)$, with entries given respectively  by the electric 
and magnetic 
charges $q ,p$, the moduli vector $G_{\a}=(g,h,l), \a=1,2,3$, the 
scalar-charge 
vector $\Sigma_{\a}=(\Sigma_{\e},\Sigma_{\s},\Sigma_{\r})$ and the ADM mass 
$M$. $\S_{\a}$ is defined in terms of the asymptotic behavior of the 
scalar fields, 
\be
2\e=2\e_{0}+{\Sigma_{\e}\over r}+O({1\over r^{2}}),\quad
2\s=2\s_{0}+{\Sigma_{\s}\over r}+O({1\over r^{2}}),\quad
2\r=2\r_{0}+{\Sigma_{\r}\over r}+O({1\over r^{2}}), 
\ee
whereas $M$ is defined, as usual, in terms of the asymptotic behavior 
of the  metric.  The previous parameters are not  independent. 
In the following, we will consider $\Sigma_\a$ and $M$ 
as a function of the independent 
parameters  $Q_{i},P_{i}$ and $G_{\a}$.

In the usual realization (see eqs. (\ref{e3}), (\ref{e4})), the duality 
group 
$O(3,Z)$ acts on the moduli
$G_{\a}$ and on the charges $Q_i,P_{i}$,$\Sigma_\a$, but it leaves the 
mass of the state 
unchanged, because the metric is left invariant. 
A spectrum-generating technic for BPS saturated states requires 
transformations at 
fixed values of the moduli. The use for this purpose of duality symmetries
is therefore problematic \cite{LPS}. In order to make $O(3,Z)$  a suitable
spectrum-generating symmetry, we use a realization of the group such 
that each
group transformation, realized as in eq (\ref{e3}), (\ref{e4}) is followed 
by the inverse 
transformation of the moduli. In this way $O(3,Z)$ acts at fixed values 
of the
moduli, whereas the mass of the state is changed. Denoting with
$G_\a\to G_{\a}'(G)$ the transformation of the moduli, we have the 
following realization of $O(3,Z)$:      
\bea \lb{e7}
G_\a &\to& G_\a, \quad Q_i\to Q_i', \quad P_i\to P_i',
\nonumber\\
\Sigma_\a&\to&\Sigma_\a'
\quad M(Q,P,G)\to M(Q,P,G'^{-1}).
\eea
Note that the transformations $G_\a\to G_\a'(G)$,  acting on the moduli 
at fixed
$U(1)$-charges, represent itself a realization of the $O(3,Z)$ group.
Their action on the spectrum is to move around in the moduli space
solutions of  given $U(1)$-charge.  
Eqs. (\ref{e7}) translate immediately in a particular realization of the 
group-generating transformations (\ref{e3}), (\ref{e4}).
In the next sections we will use this realization,  
together with
the bound state interpretation, to generate the spectrum of 
extremal black
hole solutions of the theory and to discuss their physical properties.
\section{ Elementary solutions and 
rules for the construction of bound states } 
\paragraph{}

The extremal black holes (\ref{e5}) can be viewed as bound states of 
elementary constituents \cite{RA}. It has been shown that extreme 
dilatonic 
black holes with dilaton coupling $a=1, 1/\sqrt 3, 0$ arise as bound 
states, with zero binding energy, of respectively 2,3 or 4, 
elementary, 
$a=\sqrt{3}$ black holes \cite {RA}. This idea found support 
from the fact that 
these black hole solutions can be interpreted as intersections of 
10-dimensional  D-branes, yielding after compactification to 4d the 
$a=1,1/\sqrt{3},0$ black holes \cite {PT}.  
Here, we will develop further this idea of compositeness, and we will 
use it, together with duality symmetry, to generate the  
spectrum of the BPS saturated states of the model (\ref{e1}).
Let us first list and discuss the elementary solutions, which will
be used to build  the bound states. 

We consider as elementary 
those states that have only one $U(1)$-charge different from zero. 
For fixed values of the moduli and of the U(1)-charge,  there are 
8 of such states, which we denote with $S_{i}, \bar{S_{i}}$.
They can be generated from 
\be
S_{1}: Q_{i}= (q,0,0,0),\quad \S_{\a}=qghl(-1,-1,-1),
\lb{e8}
\ee
acting with the $O(3,Z)$ group-generating transformations 
(\ref{e3}),(\ref{e4}).
We have
\bea\lb{e10}
S_{2}&:& Q_{i}= (0,q,0,0),\quad \S_{\a}=q{gh\over l}(-1,-1,1),
\nonumber\\
S_{3}&:& Q_{i}= (0,0,q,0),\quad \S_{\a}=q{g\over hl}(-1,1,1),
\nonumber\\
S_{4}&:& Q_{i}= (0,0,0,q),\quad \S_{\a}=q{gl\over h}(-1,1,-1),
\nonumber\\ 
\bar S_{1}&:& P_{i}= (q, 0,0,0),\quad \S_{\a}=q{1\over ghl}(1,1,1),
\nonumber\\
\bar S_{2}&:& P_{i}= (0,q,0,0),\quad \S_{\a}=q{l\over gh}(1,1,-1),
\nonumber\\
\bar S_{3}&:& P_{i}= (0,0,q,0),\quad \S_{\a}=q{hl\over g}(1,-1,-1),
\nonumber\\
\bar S_{4} &:& P_{i}= (0,0,0,q),\quad \S_{\a}=q{h\over gl}(1,-1,1).
\eea
The state $\bar S_{i}$ can be obtained from the state $S_{i}$ by 
acting  with the transformation $\t_{S}\t_{T}\t_{U}$. Two generic states 
verifying this relation will be called throughout this paper 
{\sl dual} states.

The charges characterize completely the state because the duality 
transformations act at fixed values of the moduli and the
mass of the single state is related to the scalar-charge 
through the simple relation
\be
M={1\over 4} |\S_{\e}|.
\lb{e12a}
\ee
The states $S_{i}$ are characterized by $M\propto g$, are electrically 
charged,
correspond to elementary string excitations \cite {DR} and are expected 
to be dominant in the weak coupling regime of the theory. 
The states $\bar S_{i}$ are characterized by $M\propto 1/g$, are 
magnetically charged ,
correspond to solitonic string excitations \cite {DR} and are expected
to be dominant in the strong coupling regime of the theory. 
This characterization holds for the $S$-String. Consistently with the 
idea of  
triality  we have besides the $S$-string also a $T$-string and a 
$U$-string, whose dilaton/axion fields are given respectively by $T$ and 
$U$ \cite{DLR}.
A state which is electric for the $S$-string can have an other 
characterization
for the other two strings. 

The information about the electric or solitonic character of  the state 
is encoded in the multiplet structure (\ref{e8}),(\ref {e10}). Of particular 
relevance is the transformation law of the states under the permutation 
group $P_3$ (this is the group that switches between the 
$S-T-U$ string description) and   the invariance group 
($\subset O(3,Z)$)  of the state.
If a state is invariant under $P_3$, then 
this state will have the same characterization in the three string 
descriptions.
This is the case of the state $S_1 (\bar S_1)$, which is electric 
(magnetic).
On the other hand, if the invariance group of the state is not $P_3$,
it will look differently for the $S-T-U$ strings and the multiplet 
structure gives account of this fact. This is what happens to  the
the other 6 elementary states in the spectrum. For example, the state 
$S_{2}$ corresponds to an electric excitation of the $S-$ and 
$T$-string but 
to a solitonic excitation of the U-string. In fact, $S_{2}$ is 
invariant under the second transformation in eq. (\ref {e4}) (in 
terms of string/string  dualities this corresponds to $H_{STU}\to 
(IIA)_{TSU}$) but is mapped into the state $\bar S_{3}$ by the 
fourth transformation  in eq. (\ref {e4}) (in 
terms of string/string dualities this corresponds to  
$H_{STU}\to (IIB)_{UTS}$).

By putting together  elementary solutions (\ref{e8}),(\ref{e10}), 
one can build  
bound  states. Starting point of this procedure are the results of ref.
\cite{RA}, where bound states of 2, 3 and 4 solutions have been
constructed. In order  to construct the whole spectrum, we need, however, 
some simple and general rules. When we form bound states of two or more 
elementary states of different $U(1)$-charges, we get a multiplicity of 
states that 
have the same non-vanishing components of the charge vectors $P$ and $Q$ 
and that differ
one from the other just in a permutation of the entries of the 
these vectors.
This is an unpleasant feature because there is no relevant physical 
information 
related to the permutation of the entries of the $U(1)$-charge vectors. 
Taking into 
account these permutations would only make our notation awkward. 
For this reason, 
we will consider as  equivalent, states that can be transformed one 
into the other
just by a permutation of the entries of the $U(1)$-charge vectors.
 We will use for this equivalence class of  composite states
(up to four elementary states) the notation,
\be
\lb{e12}
\lt S_i,S_j....\rt.
\ee
States with a given number of constituents 
transform as different multiplets under $O(3,Z)$.  We will denote 
these  multiplets 
with $n=1,2,3,4$. 

A crucial and simple feature of the composite solutions is that all
the parameters characterizing the solutions are additive, i.e. 
we can express
the charges $Q,P,\S$ and mass $M$ of the bound  state in terms of 
the charges $Q_I,P_{I},\S_I$ and 
masses $M_I$ of the elementary constituents as follows:
\be
\lb{e13}
Q= \sum Q_I, \quad P= \sum P_I,\quad \S= \sum S_I,\quad
M= \sum M_I.\quad
\ee
The additivity of the  masses implies that all the bound states 
we can construct 
have zero binding energy.
Note that the states belonging to a given equivalence class (\ref{e12}) 
in general 
are not degenerate. The charges and masses of the single states can 
be obtained 
from those 
of the state that represents the equivalence class, by permuting  the 
entries of the $U(1)$-charge vectors.

Let us now define the invariance group ${\cal G}_{0}$ of 
the state (\ref {e12}).
${\cal G}_{0}$
is defined as the subgroup of $O(3,Z)$ that maps the equivalence 
class (\ref {e12})
in itself. In general ${\cal G}_{0}$ is not the invariance group of 
the single state.
This will be given by  the product of  ${\cal G}_{0}$ with 
a subgroup of the permutation
group that acts  on the entries of the $U(1)$-charge vectors.

The previous rules enable one to write down explicitly the form of the
solution corresponding to a given bound state of elementary solutions.
However, not all possible combinations of elementary solutions are a priori
allowed. The easiest way to write down the selections rules one needs to
construct the spectrum, is to use $O(3,Z)$ duality arguments.
States  with $n= 2,3,4$ transform as  different
multiplets under the $O(3,Z)$ group and, 
owing to the duality symmetry of the spectrum, every state within 
a given multiplet  can
be generated, acting with a transformation of $O(3,Z)$, from the  
states: $(S_1,S_3)$ for $n=2$,  $(S_1,\bar S_2,S_3)$ for $n=3$,  
$(S_1,\bar S_2,
S_3,\bar S_4)$ for $n=4$.
This statement represents a strong selection rule that rules out from the
BPS spectrum some of the states that, in principle, can be build 
by combining $n=1$ elementary states. In the next section we will list
and study the allowed bound states. 
\section {Bound states}

\subsection {The $n=2$ multiplet}
By constructing all the possible combinations of  two elementary states 
we obtain  the $n=2$ bound  states.
Only 12 of these states are allowed. 
In the $S$-string description we have 2 electric states with  $M\propto g$,
\be\lb {f1}
\lt S_{1},S_{3}\rt, \lt S_{2},S_{4}\rt;
\ee
2 magnetic states with $M\propto 1/g$ ( the duals of the previous 
states),
and 8 dyonic states,
\be\lb {f2}
\lt S_{1}, \bar S_{2}\rt,
\lt \bar S_{3}, S_{4}\rt, M\propto h;
\ee
\be\lb{f3} 
\lt S_{1}, \bar S_{4}\rt,  
\lt S_{2}, \bar S_{3}\rt,   M\propto l,
\ee
together with  the dual states characterized, respectively, by 
$M\propto 1/h$ and 
$M\propto 1/l$.

From the point of view of the $T-$ and $U$-string the role of the states 
are correspondingly reversed. For instance, the $T-$string  sees the 
states (\ref{f2}) as electric, the dual states as magnetic and the states 
(\ref{f1}),  (\ref{f3}) together with their duals, as dyonic.
It is interesting to see how the existence of dyonic states is a 
consequence of the $O(3,Z)$ duality symmetry of the spectrum.
The following diagram explains the action of the $O(3,Z)$ group-generating
transformations on the $n=2$ multiplet:
\be\lb{f4} 
\begin{array}{ccccc}
(\ref {f1})&\lra& (\ref{f2})&\lra &(\ref{f3})\\
\,\,\updownarrow \t_{S}& & \,\,\updownarrow \t_{T}& &\,\,
\updownarrow \t_{U}\\
(\overline{\ref {f1}})&\lra& (\overline{\ref{f2}})
&\lra &(\overline{\ref{f3}}) 
\end{array}
\ee
The states in the second row are the dual states of those appearing 
in the first row. The horizontal  arrows represent transformations of 
the permutation group $P_{3}$ whereas the vertical arrows represent 
the $S-T-U$  dualities. The  $n=2$ multiplet is 
generated by horizontal symmetries that switch between different 
string descriptions (e.g. they  interchange states with $M\propto g $ with 
states with  $M\propto h $) together with vertical symmetries that 
map electric into magnetic states (e.g. $M\propto g$ into  
$M\propto 1/g$ states).

Each state in the multiplet has an invariance group ${\cal G}_0$ that is 
generated  by the permutation group $P_{2}$, which interchanges 
two moduli, and by the product of two dualities in (\ref{e3}), 
which act on the same moduli. For example,  the states in the first 
column of (\ref {f4}) have a invariance group generated by the 
permutation of $(T,U)$ and by  $\t_{T}\t_{U}$.
\subsection {The $n=3$ multiplet}

The $n=3$ multiplet has only 8 states. In fact the $O(3,Z)$ symmetry 
allows only for the following bound states of 3 elementary solutions:
\be\lb{f4a}
\lt S_{1},\bar S_{2},S_{3}\rt, \lt S_{1},S_{3},\bar S_{4}\rt,
\lt \bar S_{1},S_{2},S_{4}\rt, \lt  S_{2},\bar S_{3}, S_{4}\rt,
\ee
together with the dual states.
At first sight, it could seem that  these states have dyonic 
character, because they are bound  states of elementary solutions  
some of which carry magnetic, others electric, $U(1)$-charges. 
However, it turns out that we can assign to each  state either a 
electric or magnetic character. Each $n=3$ state can be put in 
correspondence with a $n=1$ state, in the following way:
\be\lb{f5} 
S_{1}\sim \lt  S_{2},\bar S_{3}, S_{4}\rt, \quad 
S_{2}\sim \lt  S_{1}, S_{3}, \bar S_{4}\rt, \quad 
S_{3}\sim \lt  \bar S_{1}, S_{2}, S_{4}\rt, \quad
S_{4}\sim \lt  S_{1},\bar S_{2}, S_{3}\rt. 
\ee
Given this correspondence, the $n=3$ multiplet spans   
the same representation 
of the $O(3,Z)$ group as the $n=1$ multiplet. As a consequence,  
the algebraic features discussed for  the $n=1$ multiplet  hold also for 
the $n=3$ multiplet. In particular,  in view of (\ref {f5}), 
the states  (\ref{f4a}) can be considered as  electric whereas   the 
dual states can be considered  as
magnetic 
($S$-string description). 

Tough the underlying $O(3,Z)$ algebraic structure 
of the $n=3$ and $n=1$  multiplets is the same, the solutions behave 
differently. As we shall see in the next section, they give rise to 
different spacetime structures and break $N=8$ supersymmetry in two 
different ways.   

\subsection {The $n=4$ multiplet}

The $n=4$ multiplet  contains only two states, the state
$X=\lt S_{1}, \bar S_{2}, S_{3}, \bar S_{4}\rt $ and its dual
$\bar X=\lt \bar S_{1}, S_{2}, \bar S_{3}, S_{4}\rt $.
These two states transform one into the other under the action of the 
$S-T-U$ dualities (\ref{e3})
but are invariant under the permutation group (\ref{e4}).
The state $X$ and $\bar X$ are, therefore, truly dyonic because all 
the three strings ($S,T$,$ U$) see them as bound states of 
two magnetic and two electric elementary states. 
This behavior is to be 
compared with that of the $n=2$ dyonic states.  The states $X$ and  $\bar X$ 
are also invariant under the action  of the  group
\be \lb{f5a}  
{\cal H}=\{ \t_{S}\t_{T},\t_{S}\t_{U}, \t_{T}\t_{U}\}.
\ee
$P_{3}$  together with ${\cal H}$ closes to form the full invariance group,
${\cal G}_{0}$, of 
the state, which is a group isomorph to $SO(3,Z)$.

\section {Spacetime structure and compositeness}

In the previous sections we have been mainly concerned with the
algebraic features of the BPS spectrum. We turn now to discuss the
spacetime structures associated with the solutions. Again, the idea 
of compositeness will play a crucial role. In particular, we will
see that there is a non-trivial interplay between the singularities, 
the topological features of the spacetime and  the 
interpretation of the solutions as bound states.

Let us first write the solution (\ref{e5}) in terms of the string metric,
the metric to which strings naturally couple. A consequence of the triality
symmetry is the existence of three string metrics, 
$g_{S},g_{T},g_{U}$,
to which the $S-T-U$ strings respectively couple. They
are  related to the canonical metric in eq. (\ref{e5}) $g_{C}$, 
through the equations
\be\lb{f5b}
g_{S}= e^{\e}g_{C}, \quad g_{T}= e^{\s}g_{C}, \quad g_{U}= e^{\r}g_{C}.
\ee
The use  of the string metric(s) for  the discussion of the spacetime 
structures
is particularly useful because it makes the transformation of the 
metric under the $O(3,Z)$ duality group  explicit. In fact, whereas the 
duality
transformations leave invariant $g_{C}$ (usual realization of the 
$O(3,Z)$ symmetry)
or change only the moduli dependence of  $g_{C}$ (our realization of the 
$O(3,Z)$ symmetry),
they  change drastically the string metrics.

 The presence of  the three string metrics 
(\ref{f5b}) 
in the same multiplet is a consequence of the $O(3,Z)$ duality 
symmetry, in particular of its $P_{3}$ subgroup. In fact, either a 
state is invariant under the action of $P_{3}$ and in this case the 
three string metrics (\ref {f5b}) are the same, or a state is not 
$P_{3}$ invariant and  in this case the multiplet must contain the
three metrics (\ref{f5b}). For this reason, it is enough to consider just 
one string metric (we choose here  $g_{S}$), the $O(3,Z)$  symmetry 
of the spectrum will take automatically into account the other two.     

In terms of the string metric, the extremal black hole solutions have  
the same form as in eqs. (\ref{e5}), but with 
metric given by
\be\lb {f6} 
ds^2=-\pu\pt \, dt^2+\lt \xd\xq\rt^{-1}(dr^2+r^2d\Omega^2).
\ee
Choosing the appropriate values of the $U(1)$-charges, and applying $O(3,Z)$ 
transformations we can generate the $n=1,2,3,4$ solutions discussed in the
previous section. We begin the discussion with some features of the general
solution.  

We are particularly interested in the near-horizon geometry, that is 
in the
solution that describes the spacetime near $r=0$.
In this region the metric (\ref{f6}) has the form
\be \lb{f7}
ds^2= -{r^m\over c^2}\, dt^2 +{b^2\over r^s}\lt dr^2+r^2 d\Omega^2\rt.
\ee
Where $m,s =0,1,2$ are, respectively,  the number of electric and 
magnetic solutions participating to the bound state, $c$ is a constant 
which depends only on the moduli  and on the electric charges, 
whereas  $b$ depends only on the moduli and on the magnetic charges.
Eq. (\ref{f7}) is an exact solution of the Weyl rescaled version of 
the model 
(\ref{e1}). This solution 
gives a general description  of the near-horizon geometry of the 
extremal black hole. The information about the nature of the state 
is encoded
in  the parameters $m,s,c,b$.
To discuss the singularities of the spacetime (\ref {f7}), which are 
the same as those of the spacetime (\ref{f6}), it is useful to write 
down the scalar curvature. It turns out that, independently of the 
parameter $m$, 
the spacetimes with $s=2$ (the bound states of 2 magnetic 
elementary constituents)  have constant, $m$-dependent, curvature
\be\lb{f8}
R= \lt 2-{m^{2}\over 2}\rt {1\over b^{2}}. 
\ee
Moreover, for $s=2$,  the spacetime (\ref{f7}) 
is 
the product  of two two-dimensional spaces of constant curvature:
\be\lb{f9}
H^{2}\times S^{2 }
\ee
where $H^{2}$ is a two-dimensional spacetime and $S^{2}$ is the two-sphere.
For $s\neq 2$ the scalar curvature is
\be\lb{g1}
R=\ha \lq m^{2}+2m -ms -4+ (2-s)^{2}\rq {r^{s-2}\over b^{2}}
\ee
In this case $r=0$ is always a curvature singularity of the  spacetime. 

\subsection {Elementary states}

The states of the $n=1$ multiplet are characterized by one non-vanishing  
value of the 
$U(1)$-charges. Therefore they give rise to two spacetime structures,
which carry either electric or  magnetic charge.  
These spacetime structures, with $h=l=1$, are also solutions of the dilaton 
gravity 
model with dilaton coupling parameter $a=\sqrt 3$. Near the horizon 
the solutions are given by eq. (\ref{f7}) with, respectively, $m=1,s=0$ and 
$m=0, s=1$,
\bea \lb{g2}
ds^2&=&g^{2}\lq  -{r\over q_{1}ghl} dt^2 +\lt dr^2+r^2 
d\Omega^2\rt\rq,\\
ds^2&=&g^{2}\lq   dt^2 + {p_{1}\over ghl}{1\over r}\lt dr^2+r^2 
d\Omega^2\rt\rq.
\eea
The previous equations describe the $S_{1}$ and the $\bar S_{1}$ state,
but with  slightly  modifications (the moduli and charge dependence) 
also the other electric ($S_{i}$) and magnetic ($\bar S_{i}$) states 
of the $n=1$ multiplet. Both the electric and magnetic spacetimes have 
a curvature singularity at $r=0$, which is null in the electric case 
and  naked (timelike) in the magnetic case.
As we shall see later in detail, the $n=1$ states are the only states 
in the spectrum that produce geometrical structures with naked 
singularities. The elementary solution have a maximal 
number ($N=4$) of unbroken 
supersymmetries. In fact,  they  saturate  the four Bogomol'nyi bounds.
This two facts are consistent with the idea of compositeness:
the elementary states are truly singular and possess a maximal amount 
of supersymmetries whereas the bound states, having an internal 
structure,  are non-singular and possess a lesser number of unbroken 
supersymmetries.

\subsection {$n=2$ bound states}
$n=2$ states can be build in three distinct ways: a) with two 
electric, b) with two magnetic and c) with one electric and one 
magnetic $U(1)$-charge.
The solutions with equal values of the charges and appropriate values of the 
moduli,
are also solutions of the $a=1$ dilaton gravity model \cite {KP}. 
Near the horizon the solutions   are, respectively,
\be\lb{g3}
ds^2=- {r^{2}\over q_{1}q_{3}} dt^2 +g^{2}\lt dr^2+r^2 
d\Omega^2\rt, \quad 
\ee
\be\lb{g4}
ds^2=- g^{2} dt^2 +{p_{1}p_{3}\over r^{2}}\lt dr^2+r^2 
d\Omega^2\rt, \quad 
\ee
\be\lb{g5}
ds^2=- {g\over h l q_{1}} r dt^2 +{p_{2} gl\over h} {1\over r}\lt dr^2+r^2 
d\Omega^2\rt. \quad 
\ee
The  equations (\ref {g3}),(\ref{g4}),(\ref{g5}) represent the (string) 
metric  
associated, respectively,  with the states $(S_{1}, S_{3}),$ $(\bar S_{1}, 
\bar S_{3}), 
(S_{1}, \bar S_{2})$. The metric solutions  associated with the other 
states in 
the $n=2$ multiplet can be easily obtained by using $O(3,Z)$ duality 
transformations. They differ from the expressions (\ref {g3}),(\ref{g4}),
(\ref{g5}) 
just in the charge and moduli dependence.
 The electric (\ref{g3}) and the dyonic (\ref{g5})  spacetime 
are both singular, but the 
singularity at $r=0$ is a null singularity. On the other hand the magnetic
spacetime (\ref{g4}) is perfectly regular, has constant curvature 
$R=2/p_{1}p_{3}$ and  has the topology (\ref{f9}) 
with $H^{2}$ given by two-dimensional Minkowski space.  These features 
have a nice interpretation. The electric solutions correspond to 
elementary  string excitation whereas the magnetic solutions 
correspond to solitonic string excitations. The dyonic solutions have a 
special status. As we have already noted in sect. 5.1, their 
existence is related  to the string triality symmetry and their 
character (electric, magnetic or dyonic) depends on the string 
picture ($S,T,U$) one is using. They can not be regular because, as  noted  
at the beginning of this section,   a bound state 
requires two elementary magnetic solutions in order to be a regular 
spacetime. All the $n=2$ BPS  states  preserve $N=2$ 
supersymmetry because 
they saturate two Bogomol'nyi bounds.

\subsection {$n=3$ bound states}
We have seen in sect. 5.2 that  the $n=3$ multiplet can be put in 
correspondence 
with the $n=1$ multiplet. For this reason, we have also in this case 
only two  spacetimes structures,  the electric one, build from two 
electric and one magnetic elementary states and the magnetic one, built 
from two 
magnetic and one electric elementary states. Both spacetime, with equal 
values of the 
$U(1)$-charges and $g=h=l=1$, are 
solutions of the $a=1/\sqrt3$   dilaton gravity model.
Near the horizon these  solutions become,
\be\lb{g6}
ds^2=- {r^{2}\over q_{1}q_{3}} dt^2 +{p_{2}gl\over hr}\lt dr^2+r^2 
d\Omega^2\rt, 
\ee
\be\lb{g7}
ds^2=- {r \over q_{1}hl} dt^2 +{p_{2}p_{4}\over r^{2}}\lt dr^2+r^2 
d\Omega^2\rt
\ee
Eqs. (\ref{g6}),(\ref{g7})  correspond respectively to the 
$(S_{1},\bar S_{2},S_{3})$ and to the $(S_{1},\bar S_{2},\bar S_{4})$ states.
The other solutions of the $n=3$ multiplet can be easily generated 
using $O(3,Z)$ duality transformations. 
They have the same structure as (\ref{g6}) and (\ref{g7}).  
As expected, the electric spacetime (\ref{g6}) has a null singularity at 
$r=0$ 
whereas the magnetic one is regular. The latter is a spacetime with constant 
curvature $R=3/(2  p_{2}p_{4})$ and of topology (\ref{f9}), with $H^{2}$ 
given by two-dimensional anti-de Sitter spacetime. Four-dimensional 
spaces of this kind have  already been studied in the context of 
four-dimensional 
string effective theories in ref. \cite {CM}.
The  $n=3$ solutions saturate only one 
Bogomol'nyi bound, they  are BPS states with  only one unbroken 
supersymmetry. 

\subsection {$n=4$ bound states}

Associated with the $n=4$  multiplet there is only one spacetime 
structure, which
appears also as solution of the $a=0$ dilaton gravity model. The uniqueness
of the solution is related to the high degree of symmetry of the $n=4$
states. In fact the two states of the $n=4$ multiplet have truly dyonic 
character and can be 
constructed  by putting together two electric and two magnetic
elementary states. The near-horizon geometry is given by the well-known
Bertotti-Robinson spacetime
\be\lb{g8}
ds^2=- {r^2 \over q_{1}q_3} dt^2 +{p_{2}p_{4}\over r^{2}}\lt dr^2+r^2 
d\Omega^2\rt.
\ee
The spacetime is again given by the product of two two-dimensional
spaces of constant curvature (\ref{f9}) ($H^2$ is also here a 
two-dimensional anti-de Sitter
spacetime). Apart from the uniqueness, this solution has other features 
that
make it peculiar with respect to the other  solitonic solutions 
discussed in this section.
First of all, the solution has the same form when expressed in terms
of the string or canonical metric. We have already noted in sect. 5.3 
that the
the $S-T-U$ string  descriptions of the $n=4$ states are the same. 
Now we  learn
 that the
string and canonical description of these states  are also the same.
Second, the scalar curvature of the spacetime (\ref{g8}) not only is constant 
but is 
actually zero; the positive curvature of the two-sphere $S^2$ compensate
exactly the negative curvature of $H^2$.
Third, the near-horizon geometry (\ref{g8}) is moduli independent and 
depends only
on the charges \cite {FK,FK1}. Last, whereas the solution (\ref{e5}) 
preserve only $N=1$ supersymmetry,
for the near-horizon geometry we have the doubling of unbroken 
supersymmetries to $N=2$
\cite{FK,KP}. These features are a consequence of the 
symmetries of the state and can be used to explain why, near the horizon, 
the $n=4$ 
solutions behave as supersymmetric attractors, whereas this is not the 
case for the 
 $n\neq4$
supersymmetric  solutions. This will be discussed  in detail in the 
next section.     

\section {Supersymmetric attractors} 
Ferrara and Kallosh have shown that in the context of $N=2$ 
supergravity theories the near-horizon geometry of a wide class of
 extreme black holes 
behaves has a supersymmetric attractor  \cite{FK,FK1}. The near-horizon 
geometry is
moduli independent and depends only on the charges. A procedure to determine 
the entropy of the extremal black hole was also given. It consists in 
taking the extrema of the  ADM mass, which for BPS saturated states 
coincides with the 
largest eigenvalue of the central charge matrix, 
with respect to the moduli at fixed $U(1)$-charges. One can show 
that the area 
of the horizon and then the entropy of the black hole, is proportional 
to $M^{2}$ evaluated in its minimum. 

A nice example of this kind of behavior is given by the $n=4$ solution 
discussed in the previous section \cite{FK}. All the features of  the 
solution fit very well in this schema. The near-horizon geometry (\ref{g8})
is independent of the moduli. The vanishing of the Weyl tensor (related to
 the vanishing of the scalar curvature) for the 
Bertotti-Robinson geometry  is responsible for the doubling of 
supersymmetries near the horizon \cite {FK,KP}. By minimizing the ADM mass 
of the solution
\be\lb{h1}
M={1\over 4}\lg q_1 ghl +q_3 {g\over hl}+p_2 {l\over gh}+
p_4 {h\over gl}\rg,
\ee
with respect to the moduli $g,h,l$ and calculating  the value
of $M$ at  the minimum one finds a non-vanishing charge-dependent,
value of the entropy.

Though the Ferrara-Kallosh mechanism works very well for the
$n=4$ states, it cannot be used (at least in its original form) 
for  $n\neq 4$. 
In fact, one easily realizes that for the $n\neq 4$ states the near-horizon
metric (both in the canonical and string description) depends on the moduli.
Moreover, the ADM mass   does not have
local extrema in the moduli space.
The existence of supersymmetric attractors and the validity of the previous 
minimization procedure does not seem  to be  features of 
generic $N=2$ BPS saturated states, but they seem to be related with 
the Bertotti-Robinson form of 
the near-horizon geometry.  
In the following, using duality symmetry arguments, we 
will show  that a minimization procedure can also be used to calculate 
the entropy of the $n\neq 4$ states and we will explain why for these  
states the near-horizon solution cannot be considered as an attractor.

It is a general fact, known since  early works on the subject,
 that dilatonic extremal 
black holes with dilaton coupling  $a\neq 0$ have zero area of the horizon
\cite {HW}, 
i.e.  zero entropy. This is   not only true for the solutions of the 
model that admit a single scalar description , but also for the generic 
solutions with $n\neq 4$. Only the states of the $n=4$ multiplet
are characterized by a non-vanishing value of the entropy.

The vanishing value of the entropy in the $n\neq 4$ case can be also
obtained by looking for the minima of the corresponding ADM masses in 
the moduli space at fixed $U(1)$-charges.
Differently from the $n=4$ case we now  look for generic, 
i.e {\sl{absolute}} 
or local, extrema
of $M$. 
Let us see how it does work in detail for the $n=1,2,3$ multiplets.

For $n=1$,  the ADM mass of the solutions can be easily calculated using eqs.
(\ref{e8}), (\ref{e10}), (\ref{e12a}). The  minima of $M$ occur at $g=0$ 
for the electric solutions and at $g=\infty$ for the 
magnetic ones. They  are characterized by finite, undetermined values of 
the other two moduli  $h,l$ and by $(M)_{min}=0$. The permutation symmetry 
of the spectrum 
implies that a pattern of $M=0$  minima can be obtained 
either with $h=0,\infty$ and  
$g,l$ undetermined  or with $l=0,\infty$ and $g,h$ undetermined. 
For example, for the state $S_{2}$ the  minimum of $M$  can be 
reached for  $g=0$ or  $h=0$ or $l=\infty$. The 
physical meaning beyond this, at first sight complicated, pattern is 
simple: given a string description, the electric 
states are 
dominant (i.e. have zero mass) in the weak coupling regime,
whereas the magnetic ones are dominant  in the strong coupling regime. 

The 12 states of th $n=2$ multiplet belong to three classes 
having  respectively 1) $M\propto g$ or $M\propto 1/g$; 
2) $M\propto h$ or $M\propto 1/h$; 3) $M\propto l$ or $M\propto 1/l$.
For instance, the states $(S_{1}, S_{3})$, $ (\bar S_{1}, 
\bar S_{3})$ have respectively 
\bea\lb {g9}
M&=&{1\over 4} g\lt q_{1}hl + {q_{3}\over hl}\rt, \\    
M&=&{1\over 4} {1\over g}\lt p_{3}hl + {p_{1}\over hl}\rt.
\eea     
The minima of $M$  fix only one modulus and 
are given respectively by 1) $g=0$ or $g=\infty$; 2) $h=0$ or $h=\infty$; 
3) $l=0$ or $l=\infty$.
As  in the $n=1$ case, the minima occur at $M=0$ and leave undetermined 
two moduli. 
However, the nature of the indetermination here is different. 
Whereas for $n=1$ the minima can occur for vanishing (or 
divergent) values of more then one modulus, for $n=2$ we 
have minima for vanishing (or divergent) values of a single modulus.
This has a simple explanation. If a state of the $n=2$ multiplet is 
electric  (or magnetic) in a string description it will be dyonic
in the other  two strings descriptions. For instance the state 
$(S_{1},S_{3})$
corresponds to elementary electric states   of the $S$-string but to 
dyonic states of the $T$- or $U$- string. Therefore it will be 
dominant in the 
weak coupling regime of the $S$-string but it will decouple in both 
the weak/strong coupling regimes of the $T$- and $U$-string. 

For  $n=3$,  we have a similar behavior as  for  $n=1$. The 
points of the moduli space which are  minima for the 
$n=3$ states are also minima for the corresponding (see eq. (\ref 
{f5}))
$n=1$ states. There is  one important difference. For $n=3$  
the indetermination that we have for  $n=1$  is removed, so that now minima 
take place on points not on curves of the moduli space. For example, 
the state $(S_{2}, \bar S_{3},S_{4})$ has mass
\be\lb{g10}
M={1\over 4} \lg g\lt {h q_{2}\over l} + {l q_{4}\over h}\rt +
{hl p_{3}\over g}\rg.
\ee
The previous function reaches the minimum at $g=h=l=0$.
The mass of the corresponding $n=1$ state, $S_{1}$,
reaches the minimum  at $g=0$, $h$ and $l$ undetermined, $h=0$, $g$ 
and  $l$ 
undetermined, $l=0$, $g$ and  $h$ undetermined.

Let us try to understand how the previous  
features  are related to the duality symmetries of the spectrum.
A crucial role  is played by the invariance group ${\cal G}_{0}$. 
We have already noted in sect. 4 that the single state belonging to 
the equivalence 
class of states (\ref{e12}) is invariant under a group  
${\cal G}_{0}\times {\cal L}_{0}$, where
${\cal L}_{0}$ is a subgroup of the permutation group acting on the 
entries of the
two $U(1)$-charge vectors $P,Q$.
It is clear that ${\cal G}_{0}\times {\cal L}_{0}$ leaves also invariant 
the ADM mass of the solution and  the near-horizon metric. 
One can show that if this group  contains as subgroup the three duality 
transformations $\tau_{S},\tau_{T},\tau_{U}$, or a group generated by 
taking products of these transformations, then {\sl a)} the minimum of $M$ 
takes 
place at finite, non-vanishing, charge-dependent 
values of the moduli and of $M$ itself; {\sl b)} the near-horizon 
geometry associated with the solution behaves as an attractor 
(it is moduli-independent and depends only on the $U(1)-$charges).
To prove the  statement {\sl a)} one has to take into account that,
as a consequence of the invariance of  $M$,  both the point of 
minimum in the moduli space  and $(M)_{min}$ must be fixed points 
of the group ${\cal G}_{0}\times {\cal L}_{0}$. Because zero and 
infinity cannot be fixed 
points of the transformations (\ref {e3}) and because the group 
${\cal  L}_0$ 
 acts on the charges $Q,P$, it follows  immediately 
the statement {\sl a)}.
Also the statement {\sl b)} follows from similar reasoning. Invariance of 
the near-horizon geometry under ${\cal G}_{0}\times {\cal L}_{0}$ implies 
that the moduli and charge dependence of the metric must be in an 
${\cal G}_{0}\times {\cal L}_{0}$-invariant combination. 
 Because the only  invariant 
of this kind that can be constructed by combining moduli and charges is 
the ADM 
mass $M$ (or a function of $M$) and because $M$ appears as the $1/r$ 
coefficient in the asymptotic expansion of the  metric, it follows that
only a charge dependence  of the near-horizon metric is allowed, i.e.  
statement {\sl b)}. 

Now we can easily  understand why the $n=4$ states are characterized 
by a non-vanishing 
value of the entropy and why for these states the near-horizon geometry 
behaves as an attractor. This two features 
simply follow  from the fact that the 
invariance group of the state   contains the  
group ${\cal H}$ in eq. (\ref{f5a}), which is generated by the dualities 
(\ref{e3}).
On the other hand, we can also explain why the states of the $n=1,2,3$ 
multiplets have 
zero entropy and why in this case the near-horizon geometry does not 
behave as an 
attractor. The invariance group of the $n=1,3$ states is $P_{3}$ and 
that of the $n=2$ states is generated by $P_2$ and by the product of 
only two of
the duality transformations (\ref {e3}) (see section 5.1). Both groups 
do not contain all the 
three dualities in eq. (\ref{e3}).

It is interesting to note that the  states of the $n=2$ multiplet 
 points out an 
intermediate behavior between the states of the $n=1,3$  and  of the  
$n=4$ 
multiplets.
Though the invariance group of the $n=2$ states does not contain all
the three duality transformations 
 of eq. (\ref{e3}), it is generated by two of them.
As a consequence though the near-horizon geometry  is 
not fully moduli independent, it depends only on one modulus.
The independence of the near-horizon geometry from two moduli means 
that, owing to the symmetries  of the state, part of the 
attractor behavior of the $n=4$ states still survives for $n=2$.  

\section {Conclusions}

One important result of our investigation of $d=4$ extreme 
black hole solutions of 
string theory is that   duality symmetries
reveal all their predictive power when used together with  
the bound state interpretation. This is at least true for the 
truncated model we have analyzed in this paper, but one expects it to 
hold for more general and realistic models. 

The  use of these two ideas  not only has given  us a powerful and 
simple tool to generate the spectrum of the BPS black holes of the theory, 
but also a key
concept to interpret the features of the black holes in terms of 
string states. Though our arguments are mostly based on the low-energy, 
truncated, string action (\ref{e1}),  the structure of the 
various multiplets reflects the underlying $O(3,Z)$ duality symmetry,
whose origin is of string theoretical nature.  In particular, using 
the string  metric to describe
the corresponding spacetime structures, we have 
found how the cosmic censor hypothesis can be implemented 
in the case when one considers black holes as bound states.
Only the elementary black hole solutions present naked
singularities. The bound states have either null singularities  
(electrically charged black holes),  or are regular geometric structure
(magnetic or truly dyonic black holes). Moreover in the latter case 
the near-horizon geometry is  the product of two two-dimensional 
spaces of constant curvature. 

Our approach has also helped us to understand some old (and new) puzzling 
features of  charged black holes in string theory. 
String theory allows for two kinds of 
extreme charged black holes.  We have extreme black holes with non-vanishing 
entropy and 
attractor behavior of the near-horizon geometry on the one side and 
extreme black holes  with 
zero entropy and  moduli-dependent near-horizon geometry on the other 
side.
We have seen that this  can be explained in terms of the duality 
symmetry of the spectrum and using the bound state interpretation.  
   
Our investigation has been mainly based on the model (\ref{e1}) that 
describes the solutions of (truncated) heterotic string theory 
compactified on a six-torus or the, through string/string dualities, 
related solutions of type $IIA, IIB$,  string theories.  It would be very 
interesting to test the validity of our approach and of our results
on the complete (not truncated) theory. In this case one has to do 
with duality groups much bigger than $O(3,Z)$  and with a huge number 
of $U(1)$ field strengths that may introduce some complications in the 
construction of the bound states.
An other interesting way to check and to improve our results is to use 
the interpretation of 4d black holes as intersections of D-branes. 
We expect this approach to be particularly fruitful because 
D-brains are  solitonic excitation of string theory, so that their use 
represent a natural framework  to implement both the ideas of duality 
and compositeness.


\begin{thebibliography}{1}

\bibitem{RD}
M.J. Duff, R.R. Khuri and J.X. Lu, \PRP  {\bf 259} (1995) 213;
C. M. Hull and P.K. Townsend, \NPB {\bf B438} (1995) 109;
E. Witten, \NPB {\bf B443} (1995) 85; 
D. Youm, {\em Black holes and solitons in string theory}, 
hep-th/9710046.

\bibitem{ST}
A. Strominger and C. Vafa, \PLB {\bf B379} (1996) 99;
G.T. Horowitz and A. Strominger, \PRL {\bf 77} (1996) 2368;
G.T. Horowitz, J. Maldacena and A. Strominger, \PLB {\bf B383} (1996) 
151.

\bibitem{DB}
J. Dai, R.G. Leigh and J. Polchinski, {\em Mod. Phys. Lett.} {\bf A4}
(1989) 2073;
J. Polchinski, \PRL {\bf 75} (1995) 4724.

\bibitem{DR}
M.J. Duff, J. Rahmfeld, 
\PLB {\bf B345} (1995) 441.

\bibitem{PT}
G. Papadopoulos and P.K. Townsend,
\PLB {\bf B380} (1996) 273;
A.A. Tseytlin, 
\NPB {\bf B475} (1996) 149;
K. Behrndt, E. Bergshoeff and J. Janssen,
\PRD {\bf D55} (1997) 3785;
V. Balasubramanian and V. Larsen,
\NPB {\bf B478} (1996) 199.

\bibitem{TS}
A.A. Tseytlin, 
\MPLA {\bf A11} (1996) 689.

\bibitem{KLOP}
R. Kallosh, A. Linde, T. Ortin, A. Peet,
\PRD {\bf D12} (1992) 5278.

\bibitem{FK}
S. Ferrara, R. Kallosh,
\PRD {\bf D 54} (1996) 1514.


\bibitem{FK1}
S. Ferrara, R. Kallosh,
\PRD {\bf D 54} (1996) 1525.


\bibitem{FKS}
S. Ferrara, R. Kallosh and A. Strominger,
\PRD {\bf D 52} (1995) 5412.

\bibitem{RA}
J. Rahmfeld,
\PLB {\bf B372} (1996) 198. 

\bibitem{MS}
J. Maharana and J. Schwarz,
\NPB {\bf B390} (1993) 3.

\bibitem{SE}
A. Sen,
\IJMP {\bf A9} (1994) 3707.

\bibitem{DLR}
M.J. Duff, J.T. Liu and J. Rahmfeld,
\NPB {\bf B459} (1996) 125.

\bibitem{LP}
H. L\"u and C. N. Pope,
\NPB {\bf B465} (1996) 127;
\IJMP {\bf A12} (1996) 437.

\bibitem{CY}
M. Cvetic and D. Youm 
\PLB {\bf B359} (1996) 87;
M. Cvetic and A.A. Tseytlin,
\PLB {\bf B366} (1996) 95; \PRD {\bf D53} (1996) 5344. 

\bibitem{OR}
T. Ortin,
{\em Extremality versus supersymmetry in stringy black holes},
hep-th/9612142.

\bibitem{LPS}
H. L\"u, C. N. Pope and K. S. Stelle,
{\em Multiplet structures of BPS solitons},
hep-th/9708109.
 
\bibitem{KP}
R. Kallosh, A. Peet,
\PRD {\bf D46} (1992) 5223.

\bibitem{CM}
M. Cadoni, S. Mignemi,
\NPB {\bf B427} (1994) 669; \PRD  {\bf D51} (1995) 4319.

\bibitem{HW}
C.F.E. Holzhey and  F. Wilczek,
\NPB {\bf B380} (1992) 447. 
\end{thebibliography}
\end{document}